\documentclass{ws-p8-50x6-00}

\begin{document}

\title{Correlations and Fluctuations at RHIC}

\author{Toru Sugitate}

\address{Physics, Hiroshima University, Higashi-Hiroshima 739-8526, Japan\\ 
E-mail: sugitate@hepl.hiroshima-u.ac.jp}


\maketitle

\abstracts{Particle correlations and fluctuations measured by 
RHIC experiments at $\sqrt{s_{NN}}$=130 GeV were discussed. 
The source size parameters were similar to those measured at 
the CERN-SPS, and no long duration time of particle emission 
were observed. It was pointed out that the dependences of 
longitudinal and transverse radius parameters on the pair 
momentum are explained with a single $m_T$ scaling function 
observed at the SPS energy. Fluctuation studies of mean $p_{T}$ 
of charged particles and of mean $E_{T}$ in an electromagnetic 
calorimeter found no significant non-statistical fluctuations 
by PHENIX, but some indication in charge independent $<p_T>$ 
by STAR.}

\section{Introduction}
The correlation and fluctuation studies of hadrons in 
ultra-relativistic heavy-ion collisions reveal particle dynamics 
in a hadronization process at a thermal freeze-out temperature, 
where strong interactions between constituent particles in hadronic 
gas end substantially as a result of expansion of the volume. 
The particle yields and their momentum distributions including 
correlation and fluctuation signals are fixed at this stage. 
The particle correlations tell us the space-time evolution of 
the hadronic gas, while the particle fluctuations provide 
information of diffusion dynamics of the fluid. It has been 
discussed that such hadronic gas might be formed via a 
Quark-Gluon-Plasma(QGP) phase after the quark fluid cools down, 
if QGP matter is created during a collision. In such a scenario, 
signals of particle correlations and fluctuations could contain 
remnants of another multi-particle dynamics of de-confined quarks 
and gluons in the QGP epoch.

\section{Correlation measurements}
Suppose matter made of hadrons has an alternative phase of 
matter composed of de-confined quarks and gluons. Such the 
partonic matter should indicate different relations among 
thermo-dynamical variables from that of the hadronic matter 
since of a larger degree of freedom and due to presence of 
the Bag pressure in the hadronic phase. If QGP matter is 
created during a nuclear collision, the matter has to transit 
to normal nuclear matter beyond an energy gap at a critical 
temperature or density, since we have not seen any partonic 
matter at zero temperature in this world. The QGP matter 
explosively expands as cooling down until the body reaches 
the critical temperature or density. In case the phase 
transition is in the first order, the expansion may slow down 
at the phase boundary due to a softening of the equation of 
state. Consequently a prolonged lifetime of particle emission 
is anticipated. The duration time of particle emission is an 
experimentally observable quantity in particle correlation 
measurements as a difference between the widths of the peaks 
in the correlation functions in the direction of pair (``outward'') 
and perpendicular to it (``sideward'') in the longitudinal 
center of mass system (LCMS).~\cite{Bertsch89} Therefore a 
larger source size parameter in the outward direction than 
that in the sideward direction can be a signature of QGP formation. 
Certainly we have to note that this idea should work completely 
for a static particle source, but the particle source in question 
is explosively expanding.

The STAR collaboration~\cite{STARPRL87} has published the first 
RHIC result of two-pion correlations measured at the mid-rapidity 
in the 12\% most central event class in Au+Au collisions. The 
PHENIX collaboration~\cite{PHENIXQM01} has also presented their 
pion correlation data in the same collision. The data are quite 
consistent each other. The systematic study of source size 
parameters as a function of c.m. energy indicates no jumps or 
steep changes of the parameters from the SPS to RHIC energies, 
but shows a gradual decrease of chaoticity and an increase of 
the longitudinal source size as is expected in the boost 
invariant model. 

\begin{figure}[t]
\epsfxsize=30pc 
\epsfbox{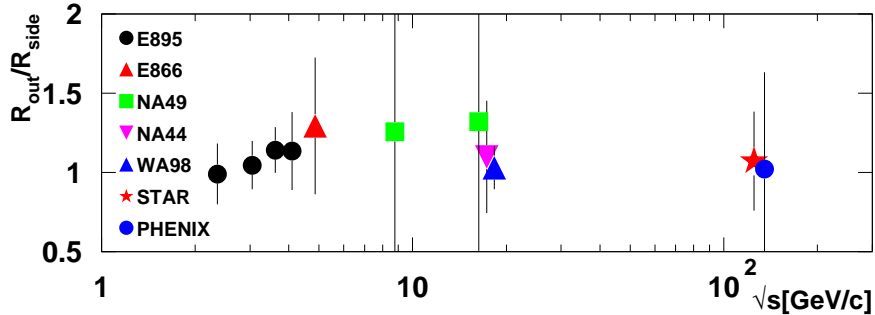} 
\caption{The ratio of $R_{out}$ to $R_{side}$ in the LCMS at various c.m. 
energies from AGS to RHIC.} 
\label{fig:Ratio}
\end{figure}

The ratios of $R_{out}$ to $R_{side}$ at various energies from 
AGS to RHIC are compiled in Fig.\ref{fig:Ratio}. There were 
predictions that the ratio could be double or larger at the RHIC 
energy, but they are excluded in the Au on Au collisions at this 
energy. The ratios seem consistent with unity over the entire 
range, and we conclude that no large temporal components are 
observed in the $R_{out}$ measurement. 

\begin{figure}[t]
\epsfxsize=30pc 
\epsfbox{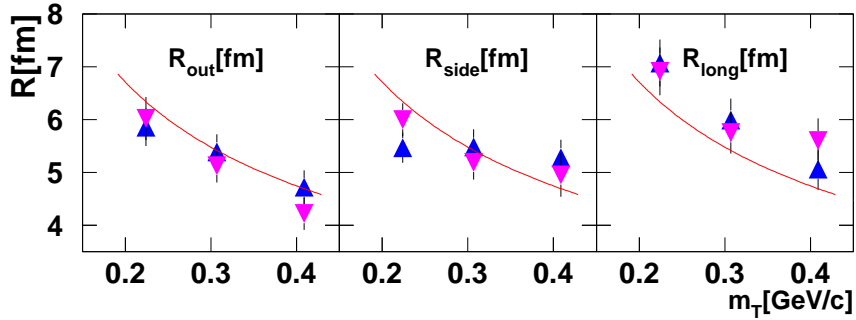} 
\caption{The source size parameters as a function of transverse mass $m_T$ 
of the pair particles observed. The lines show the $m_T$ scaling function 
observed at the SPS energy.} 
\label{fig:STARmTDependence}
\end{figure}

Figure \ref{fig:STARmTDependence} shows the source size parameters 
of the STAR data~\cite{STARPRL87} as a function of transverse mass 
of the pair particles. It indicates that positive and negative 
pions behave identical and that the $m_T$ dependence exists. 
The $m_T$ dependence in the longitudinal direction has been predicted 
by several models~\cite{Akkelin95,Lorstad96} including hydro-dynamical 
expansion in a source. Recently the NA44 collaboration~\cite{ToruPRL} 
discussed that both the transverse and longitudinal radius parameters 
at the SPS energy are fitted well with a single scaling function 
$R=A/\sqrt{m_{T}}$, where $A=3.0\pm 0.2~{\rm fm {GeV}^{1/2}}$. 
The lines inserted in Fig.\ref{fig:STARmTDependence} show the 
function, which seems to reproduce the dependences of both transverse 
and longitudinal radius parameters at the RHIC energy. We will 
examine this hypothesis in the second RHIC-year runs 
at $\sqrt{s_{NN}}$=200 GeV.

\section{Fluctuation measurements}
Fluctuation study is another tool to pin down the phase transition 
from QGP matter to hadronic matter. If a QCD mixed state where 
partonic bubbles coexist in hadronic matter, is produced in a period 
of transition, and if the transition is too fast to maintain local 
equilibrium, fluctuations not following the statistics of hadronic 
particles are anticipated in physical quantities relating to particle 
productions. Therefore, a particle fluctuation beyond statistics 
might be a signature of the QGP formation. Theorists suggest to 
measure locally conserved quantities such as net baryon number, 
electric charge or strangeness.~\cite{AsakawaQM01} 
Enormous efforts are under going to extract these quantities. 
Data available at this moment are the event-by-event mean momentum 
fluctuation of charged particles and the fluctuation of mean 
transverse energy measured in an electromagnetic calorimeter. 
The PHENIX collaboration~\cite{JeffACS} has presented the mean 
$p_T$ fluctuation of the top 5\% centrality events in Au+Au 
collisions. The distribution is quite well reproduced over four 
orders of magnitude by a Gamma function with given parameters 
extracted from the semi-inclusive $p_T$ measurements. The 
collaboration has made a very careful mathematical analysis 
and preliminarily concluded that no significant excess from 
the statistically independent emissions were observed. 
The same conclusion has been drawn for the mean $E_T$ 
distribution. 

\section{Summary}
We have discussed the new results of particle correlation and
fluctuation studies from the RHIC experiments. The source size 
parameters observed in Au on Au collisions at $\sqrt{s_{NN}}$=130 
GeV  are similar to those at the SPS energies. There are no 
indications of a huge source formation or a prolonged mixed 
phase as the similar observations at the SPS or AGS energies. 
The source size parameter dependences on the pair momentum is 
reproduced by a single $m_T$ scaling function observed at the 
SPS energy. No significant non-statistical fluctuations in $<p_T>$ 
and $<E_T>$ were observed by PHENIX, but some indication in 
charge independent $<p_T>$ is reported by STAR~\cite{STARQM01}.


\begin{thebibliography}{99}
\bibitem{Bertsch89}G. Bertsch and G.E. Brown, {\it Phys. Rev. C}
{\bf 40} 1830 (1989).

\bibitem{STARPRL87}STAR Collboration, C. Adler {\it et al}, \Journal{\PRL}
{87}{082301}{2001}.

\bibitem{PHENIXQM01}S.C. Johnson for the PHENIX Collaboration, 
talk at ``Quark Matter 2001'', nucl-ex/0104020.

\bibitem{Akkelin95}S.~V.~Akkelin and Yu.~M.~Sinyukov, 
\Journal{\PLB}{356}{525}{1995}.

\bibitem{Lorstad96}T.~Cs\"{o}rg\H{o} and B.~L\"{o}rstad, 
{\it Phys. Rev. C}{\bf 54} 1390 (1996). 

\bibitem{ToruPRL}I.G. Bearden {\it et al}, 
\Journal{\PRL}{87}{112301}{2001}.

\bibitem{AsakawaQM01}M. Asakawa {\it et al}, \Journal{\PRL}{85}{2072}{2000}.

\bibitem{JeffACS}J.T. Mitchell for the PHENIX Collaboration, 
talk at ``APS/JPS DNP Meeting in Maui'' in 2001.

\bibitem{STARQM01}J.G. Reid for the STAR Collboration, 
talk at ``Quark Matter 2001''. 

\end{thebibliography}
\end{document}